\documentclass[floatfix,aps,showpacs,nofootinbib,amssymb,amsmath]{revtex4}

\usepackage{graphicx}

\begin{document}

\title{Latest inflation model constraints from cosmic microwave background
measurements}
\author{William H.\ Kinney} \email{whkinney@buffalo.edu}
\affiliation{Department of Physics, University at Buffalo,
        the State University of New York, Buffalo, NY 14260-1500}
\author{Edward W.\ Kolb} \email{Rocky.Kolb@uchicago.edu}
\affiliation{Department of Astronomy and Astrophysics, Enrico Fermi Institute,
	and Kavli Institute for Cosmological Physics,
       	University of Chicago, Chicago, Illinois \ 60637-1433}
\author{Alessandro Melchiorri} \email{alessandro.melchiorri@roma1.infn.it}
\affiliation{Dipartimento di Fisica and Sezione INFN,
Universita' di Roma ``La Sapienza'', Ple Aldo Moro 2, 00185, Italy}
\author{Antonio Riotto} \email{antonio.riotto@pd.infn.it}
\affiliation{CERN, Theory Division, Geneva 23, CH-1211, Switzerland}
\affiliation{INFN, Sezione di Padova, Via Marzolo 8, I-35131, Padova, Italy}

\date{\today}

\begin{abstract}

We present an update of cosmological constraints on single-field inflation in
light of the Wilkinson Microwave Anisotropy Probe satellite mission five-year
results (WMAP5). We find that the cosmic microwave background data are quite
consistent with a Harrison-Zel'dovich primordial spectrum with no running and
zero tensor amplitude.  We find that the three main conclusions of our
analysis of the WMAP three-year data (WMAP3) are consistent with the WMAP5
data: (1) the Harrison--Zel'dovich model is within the $95\%$ confidence level
contours; (2) there is no evidence for running of the spectral index of scalar
perturbations; (3) From the WMAP 5 data alone, potentials of the form $V
\propto \phi^p$ are consistent with the data for $p = 2$, and are ruled out
for $p = 4$. Furthermore, consistent with our WMAP3 analysis, we find no
evidence for primordial tensor perturbations, this time with a $95\%$
confidence upper limit of $r < 0.4$ for the WMAP5 data alone, and $r < 0.35$
for the WMAP5 data taken in combination with the Arcminute Cosmology Bolometer
Array (ACBAR).

\end{abstract}


\maketitle

%%%%%%%%%%%%%%%%%%%%%%%%%%%%%%%%%%%%%%%%%%%%%%%%%%%%%%%%%%%%%%%%%%%%%%%%%
%%%%%%%%%%%%%%%%%%%%%%%%%%%%%%%%%%%%%%%%%%%%%%%%%%%%%%%%%%%%%%%%%%%%%%%%%
\section{Introduction}
%%%%%%%%%%%%%%%%%%%%%%%%%%%%%%%%%%%%%%%%%%%%%%%%%%%%%%%%%%%%%%%%%%%%%%%%%
%%%%%%%%%%%%%%%%%%%%%%%%%%%%%%%%%%%%%%%%%%%%%%%%%%%%%%%%%%%%%%%%%%%%%%%%%

Inflation \cite{lrreview} is the dominant paradigm for understanding the
initial conditions for structure formation and for Cosmic Microwave Background
(CMB) temperature anisotropies. In the inflationary scenario, primordial
density and gravitational-wave fluctuations are created from quantum
fluctuations, ``redshifted'' beyond the horizon during an early period of
superluminal expansion of the universe, then ``frozen''
\cite{Starobinsky:1979ty,muk81,bardeen83}.  Perturbations at the surface of
last scattering are observable as temperature anisotropies in the CMB, as
first detected by the Cosmic Background Explorer satellite. The latest
confirmation of the inflationary paradigm has been recently provided by the
five-year data from the Wilkinson Microwave Anisotropy Probe (WMAP) satellite
\cite{wmap5cosm,wmap5source,wmap5like,wmap5ang,wmap5fore,wmap5data}.   The
WMAP collaboration has produced new full-sky temperature and polarization 
maps in five frequency bands from 23 to 94 GHz based on the first five years
of the WMAP sky survey.  The five-year maps incorporate several improvements
in  data processing made possible by the additional years of data  and by a
more complete  analysis of the instrument calibration and in-flight beam
response. WMAP data support the inflationary model as the mechanism for the
generation of super-horizon curvature fluctuations.

The goal of this paper is to make use of the recent WMAP five-year data to
discriminate among the various single-field inflationary models. As such, this
paper represents an update of our previous analyses \cite{wmapping1,wmapping2}
of the first-year and three-year WMAP data,  respectively. The reader is
referred to Ref.\ \cite{wmapping2} for a more detailed discussion of the
inflationary model space, and for more extensive references. 

We compare our results to those reported by the WMAP team in Komatsu 
\textit{et al.}, \cite{wmap5cosm} and in Dunkley \textit{et al.}\
\cite{wmap5like}.  As first pointed out in Ref.\ \cite{dodelson97}, for
single-field inflation models, the relevant parameter space for distinguishing
among models is defined by the scalar spectral index $n$, the ratio of tensor
to scalar fluctuations $r$, and the running of the scalar spectral index $d n /
d\ln{k}$.  The Harrison-Zeldovich (H-Z) point  ($n=1,r=0$)  is comfortably
within the $95\%$ C.L.\ contour of Ref.\ \cite{wmap5cosm}, and just at the
borderline of the $95\%$ C.L.\ contour of Ref.\ \cite{wmap5like}. This small
difference appears to be statistically insignificant. In our analysis using the
WMAP5 data with slightly different priors than used by the WMAP team (discussed
in Sec.\ III) we find that marginalizing over all parameters except $n$ and $r$
in the WMAP5 data, the likelihood of the H-Z point $r = 0$, $n = 1$ is 0.0988,
corresponding to exclusion of the H-Z model with confidence level of $93.8\%$.
Marginalizing over all parameters except $n$ (\textit{i.e.}, the
single-parameter confidence limits on $n$ for the same parameter set), the
likelihood of the point $n = 1$ is 0.4226, corresponding to exclusion of the
scale-invariant spectrum with a confidence level of $81.1\%$.  These figures are highly dependent on our choice of priors: if we use a six-parameter fit with a prior of $r=0$, the likelihood of the point $n=1$ is $0.07392$, corresponding to the exclusion of the scale-invariant spectrum with a confidence level of $97.8\%$.
(In all cases, the maximum likelihood is normalized to unity.)  Our conclusion is that the H-Z
model is not preferred, but neither is it strongly disfavored.

The paper is organized as follows: In Sec.\ II we will quickly review
single-field inflation models and their observables. In Sec.\ III we define
the inflationary model space as a function of the slow-roll parameters
$\epsilon$ and $\eta$. In Sec.\ IV we describe our analysis method as well as
our results. Since a study of the implications of the WMAP5 data for single
field models of inflation has been already performed by the WMAP collaboration
themselves \cite{wmap5cosm}, we will also specify  the differences between our
analysis and theirs.  In Sec.\ V we present our conclusions.

%%%%%%%%%%%%%%%%%%%%%%%%%%%%%%%%%%%%%%%%%%%%%%%%%%%%%%%%%%%%%%%%%%%%%%%%%
%%%%%%%%%%%%%%%%%%%%%%%%%%%%%%%%%%%%%%%%%%%%%%%%%%%%%%%%%%%%%%%%%%%%%%%%%
\section{Single-field inflation and the inflationary observables}
%%%%%%%%%%%%%%%%%%%%%%%%%%%%%%%%%%%%%%%%%%%%%%%%%%%%%%%%%%%%%%%%%%%%%%%%%
%%%%%%%%%%%%%%%%%%%%%%%%%%%%%%%%%%%%%%%%%%%%%%%%%%%%%%%%%%%%%%%%%%%%%%%%%

Inflation not only explains the large-scale homogeneity of the universe, but
also provide a mechanism for explaining the observed level of
\textit{inhomogeneity} as well. During inflation, quantum fluctuations on small
scales are quickly redshifted to scales much larger than the horizon size,
where they are ``frozen'' as perturbations in the background metric. The metric
perturbations created during inflation are of two types: scalar, or
\textit{curvature} perturbations, which couple to the stress-energy of matter
in the universe and form the ``seeds'' for structure formation, and tensor, or
gravitational-wave perturbations, which do not couple to matter.  Both scalar
and tensor perturbations contribute to CMB anisotropies. Scalar fluctuations can
also be interpreted as fluctuations in the density of the matter in the
universe. Scalar fluctuations can be quantitatively characterized by the
comoving curvature perturbation $P_{\cal R}$. As long as  slow-roll is
attained, the curvature (scalar) perturbation at horizon crossing  can be 
shown to be \cite{lrreview} 
\begin{equation} 
P_{\cal R}^{1/2}\left(k\right) =
\left(\frac{H^2}{2 \pi \dot \phi}\right)_{k = a H} =    
\left [\frac{H}{m_{\rm Pl} } \frac{1}{\sqrt{\pi \epsilon}}\right]_{k = a H}. 
\end{equation} 
Here, $\phi$ denotes the slow-rolling scalar field dominating the energy
density of the Universe during inflation, the so-called \textit{inflaton},  $H$
is the Hubble rate  and $m_{\rm Pl}=1.2\times 10^{19}$ GeV is the Planck scale.
The slow roll approximation is consistent if both the slope and curvature of
the inflaton  potential $V(\phi)$ are small (in Planckian units),  $V',\ V''
\ll V$. In this case the slow-roll parameter $\epsilon$ can be expressed in
terms of the potential as 
\begin{equation} 
\epsilon \equiv \frac{m_{\rm Pl}^2}{4 \pi} 
\left(\frac{H'\left(\phi\right)}{H\left(\phi\right)}\right)^2 \simeq
\frac{m_{\rm Pl}^2}{16 \pi} 
\left(\frac{V'\left(\phi\right)}{V\left(\phi\right)}\right)^2. 
\end{equation} 
We will also need in the following  a second ``slow-roll parameter'' $\eta$
defined by
\begin{equation} 
\eta\left(\phi\right) \equiv \frac{m_{\rm Pl}^2}{4 \pi} 
\left(\frac{H''\left(\phi\right)}{H\left(\phi\right)}\right)
\simeq  \frac{m_{\rm Pl}^2}{8 \pi} \left[\frac{V''\left(\phi\right)}
{V\left(\phi\right)} - \frac{1}{2} \left(\frac{V'\left(\phi\right)}
{V\left(\phi\right)}\right)^2\right]. 
\end{equation} 
Slow roll is then a consistent approximation for $\epsilon,\ \eta \ll 1$. 

The fluctuation power spectrum is, in general, a function of wavenumber $k$,
and is evaluated when a given mode crosses outside the horizon during
inflation, $k = a H$. Outside the horizon, modes do not evolve, so the
amplitude of the mode when it crosses back \textit{inside} the horizon during
a later radiation- or matter-dominated epoch is just its value when it left
the horizon during inflation.  Instead of specifying the fluctuation amplitude
directly as a function of $k$, it is convenient to specify it as a function of
the number of \textit{e}-folds $N$ before the end of inflation at which a mode
crossed outside the horizon. 

The \textit{scalar spectral index} $n$ for $P_{\cal R}$ is defined by
\begin{equation}
n - 1 \equiv \frac{d\ln P_{\cal R}}{d\ln k},
\end{equation}
so that a scale-invariant spectrum, in which modes have constant amplitude at
horizon crossing, is characterized by $n = 1$. 

The power spectrum of tensor fluctuation modes and the corresponding tensor 
spectral index is given by \cite{lrreview}
\begin{eqnarray}
P_T^{1/2}\left(k_N\right) & = & \left[\frac{4 H}{m_{\rm Pl} \sqrt{\pi}}
\right]_{k=aH} \nonumber \\
n_T & \equiv & \frac{d\ln P_T}{d\ln k}.
\end{eqnarray}

The ratio of tensor-to-scalar modes is then $ P_T/P_{\cal R} = 16 \epsilon$,
so that tensor modes are negligible for $\epsilon \ll 1$. In the limit of slow
roll, the spectral indices $n$ and $n_{T}$ vary slowly or not at all with
scale.  We can write the spectral indices $n$ and $n_{T}$ to lowest order in
terms of the slow roll parameters $\epsilon$ and $\eta$ as
\begin{eqnarray}
n & \simeq & 1 - 4 \epsilon + 2 \eta,\nonumber \\
n_{T} & \simeq& - 2 \epsilon.
\end{eqnarray}
The tensor/scalar ratio is frequently expressed as a quantity $r$, which is 
conventionally normalized as
\begin{equation}
r \equiv 16 \epsilon = \frac{P_{\rm T}}{P_{\cal R}} .
\end{equation}
The tensor spectral index is \textit{not} an independent parameter, but is
proportional to the tensor/scalar ratio, given to lowest order in slow roll by
$ n_{T} \simeq - 2 \epsilon = - r/8$.  This is known as the consistency
relation for inflation.  A given inflation model can therefore be described to
lowest order in slow roll by three independent parameters: $P_{\cal R}$,
$P_{T}$, and $n$. If we wish to include higher-order effects, we would have 
to include a fourth parameter describing the running of the scalar spectral 
index, $d n / d\ln{k}$.

%%%%%%%%%%%%%%%%%%%%%%%%%%%%%%%%%%%%%%%%%%%%%%%%%%%%%%%%%%%%%%%%%%%%%%%%%
%%%%%%%%%%%%%%%%%%%%%%%%%%%%%%%%%%%%%%%%%%%%%%%%%%%%%%%%%%%%%%%%%%%%%%%%%
\section{Analysis and Results}
\label{secCMBanalysis}
%%%%%%%%%%%%%%%%%%%%%%%%%%%%%%%%%%%%%%%%%%%%%%%%%%%%%%%%%%%%%%%%%%%%%%%%%
%%%%%%%%%%%%%%%%%%%%%%%%%%%%%%%%%%%%%%%%%%%%%%%%%%%%%%%%%%%%%%%%%%%%%%%%%

The method we adopt is based on the publicly available Markov Chain Monte Carlo
(MCMC) package \texttt{cosmomc} \cite{Lewis:2002ah}. We sample the following
eight-dimensional set of cosmological parameters, adopting flat priors on them:
the physical baryon and CDM densities, $\omega_b=\Omega_bh^2$ and
$\omega_c=\Omega_ch^2$, the ratio of the sound horizon to the angular diameter
distance at decoupling, $\theta_s$, the scalar spectral index, its running, and
the overall normalization of the spectrum, $n$, $dn/d{\rm ln}\,k$ and $A$ at
some pivot scale $k$ (either $k=0.002$ Mpc$^{-1}$ or $k=0.017$ Mpc$^{-1}$), the
tensor contribution $r$, and, finally, the optical depth to reionization,
$\tau$. Furthermore, we consider purely adiabatic initial conditions, we impose
spatial flatness, and we use the inflation consistency relation to fix the
value of the tensor spectral index $n_T$. We also restrict our analysis to the
case of three massless neutrino families; introducing a neutrino mass in
agreement with current neutrino oscillation data does not change our results in
a significant way.

We include the five-year data \cite{wmap5cosm} (temperature and polarization)
with the routine for computing the likelihood supplied by the WMAP team and
available at the \texttt{LAMBDA} web site http://lambda.gsfc.nasa.gov/. We
also perform an analysis based on the WMAP five-year data in combination with
results from the Arcminute Cosmology Bolometer Array (ACBAR)
\cite{Reichardt:2008ay}.  The MCMC convergence diagnostic is done on $8$
chains though the Gelman and Rubin ``variance of chain mean''$/$``mean of
chain variances'' $R$ statistic for each parameter. Our 1D and 2D constraints
are obtained after marginalization over the remaining ``nuisance'' parameters,
again using the programs included in the \texttt{cosmomc} package.
Furthermore, we make use of the HST measurement of the Hubble parameter $H_0 =
100h \text{ km s}^{-1} \text{Mpc}^{-1}$ \cite{hst} by multiplying the
likelihood by a Gaussian likelihood function centered around $h=0.72$ and with
a standard deviation $\sigma = 0.08$. Finally, we include a top-hat prior on
the age of the universe: $10 < t_0 < 20$ Gyrs. We perform the analysis in two
ways: first, we constrain parameters with a prior assumption of $dn/d\ln{k} =
0$, \textit{i.e.,} an exact power-law perturbation spectrum, and second, with
$dn/d\ln{k}$ a free parameter in the MCMC. We include lensing in the calculation, and marginalize over the Sunyaev-Zel'dovich amplitude using the WMAP V-band template.

All fits are performed at a pivot scale of $k = 0.017$, which is different
from our previous analysis and the analysis of Dunkley \textit{et al.} and
Komatsu \textit{et al.}, all of which are normalized to a pivot scale of $k =
0.002$. The reason for the different choice of pivot scale is to mitigate the
parameter degeneracy between $n$ and $dn/d\ln{k}$ \cite{Cortes:2007ak}, which
significantly improves convergence of the MCMC code. In the case of
$dn/d\ln{k} = 0$, the choice of pivot scale has no effect on the spectral
index $n$, since a change of pivot affects only the normalization:
\begin{equation}
P(k) = P_{0.002} \left(\frac{k}{k=0.002}\right)^{n - 1} 
= P_{0.017} \left(\frac{k}{k=0.017}\right)^{n - 1}.
\end{equation}
However, the change of pivot results in a rescaling of the tensor/scalar ratio
$r$ which is dependent on $r$ and $n$, since
\begin{equation}
r\left(k\right) = \frac{P_{\rm T}\left(k\right)}{P_{\cal R}\left(k\right)}
\propto k^{n_T + \left(1 - n\right)},
\end{equation}
and therefore, using the consistency relation $n_T = -r/8$,
\begin{equation}
\label{eq:pivottranslation}
r_{0.017} = \frac{P_{\rm T}}{P_{\cal R}}\bigg\vert_{k=0.017} = r_{0.002}
\left(\frac{0.002}{0.017}\right)^{n-1+r_{0.002}/8}. 
\end{equation}

\begin{figure} 
\includegraphics[width=6.5in]{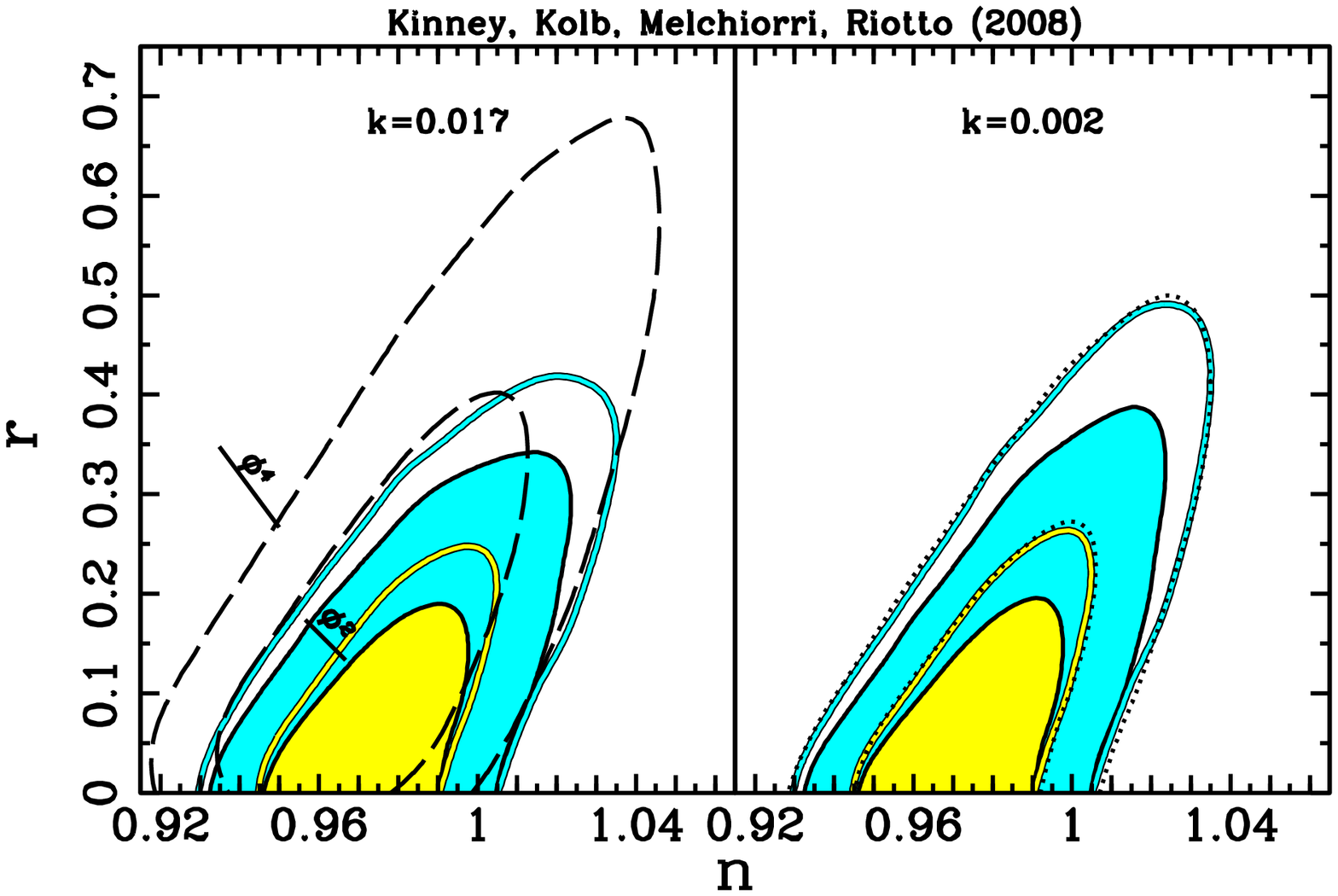}
\caption{\label{fig:WMAPNorun} 68\% C.L.\ and 95\% C.L.\ contours in the $n$,
$r$ parameter space for WMAP5 alone  (open contours) and WMAP5 + ACBAR (filled
contours), with a prior of  $dn/d\ln{k} = 0$. In the left figure the line
segments show the predictions for $V(\phi) =  m^2 \phi^2$ (lower line segment)
and $V(\phi) = \lambda \phi^4$ (upper line segment) for the number $N$ of 
\textit{e}-folds  before the end of inflation at which a mode crossed outside
the horizon in the range $N = [46,60]$. Also in the left figure the dashed 
lines show the 68\% C.L.\ and 95\% C.L.\ contours from our previous analysis 
of the WMAP 3-year data. The figure on the left is for a pivot scale of $k =
0.017$ and the figure on the right is the same contours translated to a pivot
scale of $k = 0.002$ by Eq.\ (\ref{eq:pivottranslation}). The dotted line in
the right panel is the result of a MCMC analysis run directly at a pivot of $k
= 0.002$ using the WMAP5 data alone.}
\end{figure}

Figure \ref{fig:WMAPNorun} shows $68$\% and $95$\% confidence limits on $r$ and
$n$ in the case of a no-running prior, $dn/d\ln{k} = 0$. Clearly the choice of
the pivot scale makes very little difference in this case (and since it is just
a normalization for $r$, no difference at all in the $r=0$ case.)

\begin{figure} 
\includegraphics[width=4in]{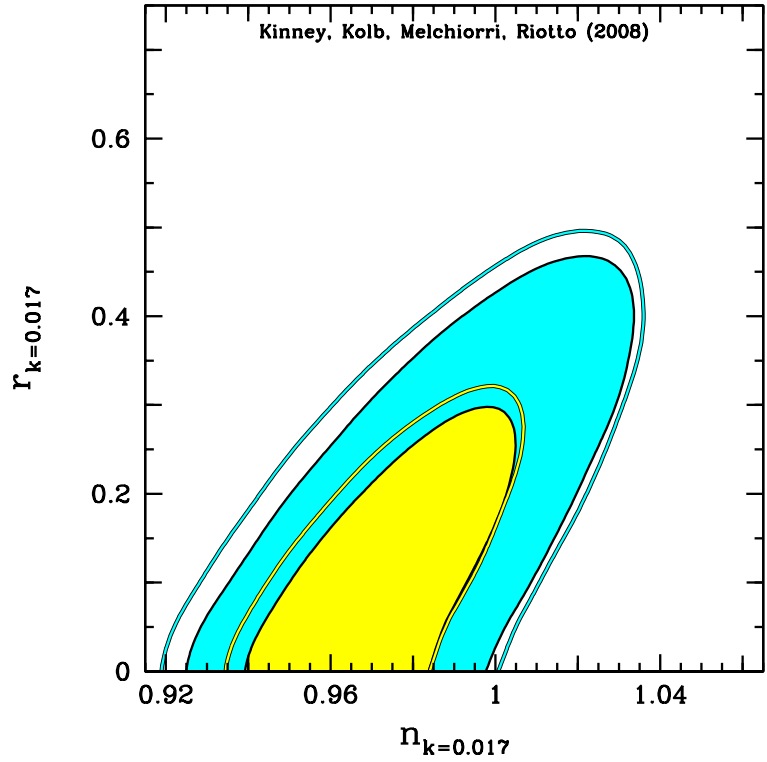}
\caption{\label{fig:run_nr} $68$\% and $95$\% confidence limits on 
$r$ and $n$ for WMAP5 alone  (open contours) and WMAP5 + ACBAR (filled
contours) allowing the possibility of a running spectral index.}
\end{figure}

In Fig.\ \ref{fig:run_nr} we show the $68$\% and $95$\% confidence limits on 
$r$ and $n$ in the case where we allow running of the spectrum $dn/d\ln{k}$ as a free parameter.  Again the WMAP5
alone data correspond to the open contours and WMAP5 + ACBAR to the filled 
contours.

Finally in Fig.\ \ref{fig:run_ndn} we show the $68$\% and $95$\% confidence
limits on $dn/d\ln{k}$ as a function of $n$ and $r$ as a function of
$dn/d\ln{k}$.

%%%%%%%%%%%%%%%%%%%%%%%%%%%%%%%%%%%%%%%%%%%%%%%%%%%%%%%%%%%%%%%%%%%%%%%%%
%%%%%%%%%%%%%%%%%%%%%%%%%%%%%%%%%%%%%%%%%%%%%%%%%%%%%%%%%%%%%%%%%%%%%%%%%
\section{Conclusions}
%%%%%%%%%%%%%%%%%%%%%%%%%%%%%%%%%%%%%%%%%%%%%%%%%%%%%%%%%%%%%%%%%%%%%%%%%
%%%%%%%%%%%%%%%%%%%%%%%%%%%%%%%%%%%%%%%%%%%%%%%%%%%%%%%%%%%%%%%%%%%%%%%%%

\begin{figure} 
\includegraphics[width=3.25in]{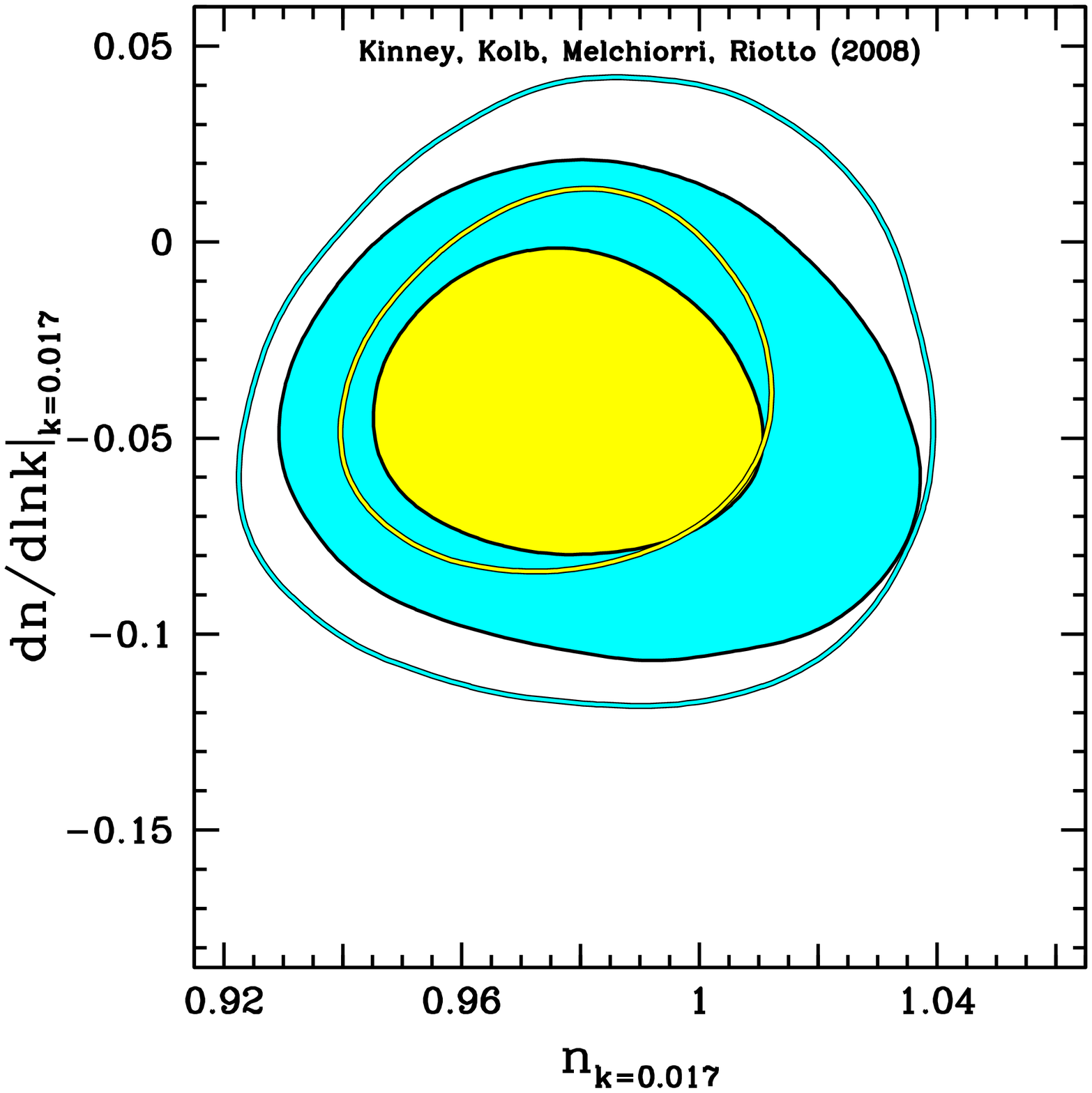}
\includegraphics[width=3.25in]{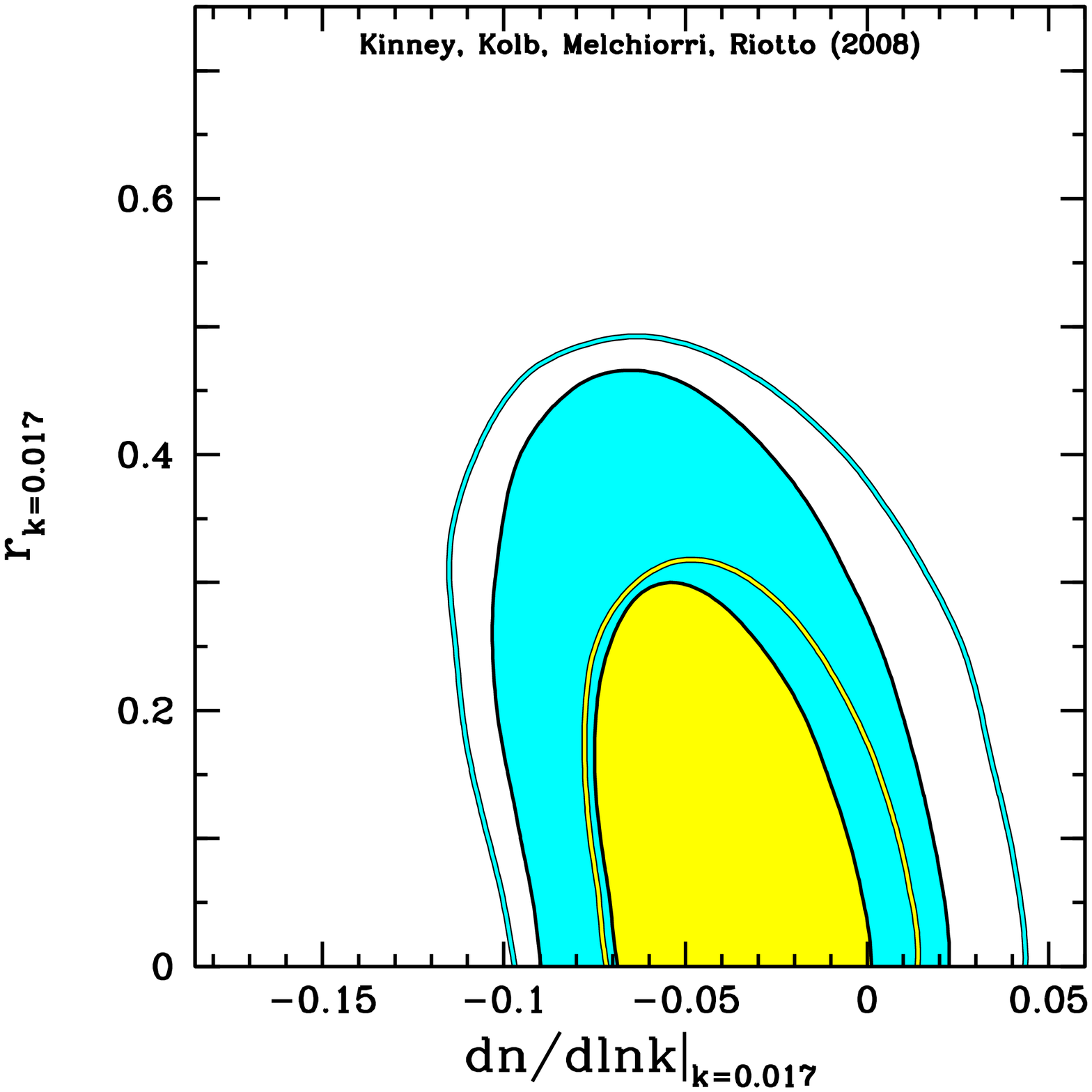}
\caption{\label{fig:run_ndn} $68$\% and $95$\% confidence limits for 
$dn/d\ln{k}$ vs.\ $n$ and $r$ vs.\ $dn/d\ln{k}$.  The different contours are 
for different data sets as in Figs.\ 1 and 2.}
\end{figure}

In this paper we presented an analysis of the recent WMAP five-year data set
with an emphasis on parameters relevant for distinguishing among the various
possible models for inflation. This analysis is an update to our similar
analysis of the WMAP three-year data set \cite{wmapping2}. In that analysis, we
reached three main conclusions, which are either unchanged or slightly
strengthened by the new data: (1) the Harrison--Zel'dovich model is within the
$95\%$ confidence level contour.  We find that marginalizing over all
parameters except $n$ and $r$ in the WMAP5 data, the likelihood of the H-Z
point $r = 0$, $n = 1$ is 0.0988, corresponding to exclusion of the H-Z model
with confidence level of $93.8\%$. Marginalizing over all parameters except $n$
(\textit{i.e.}, the single-parameter confidence limits on $n$ for the same
parameter set), the likelihood of the point $n = 1$ is $0.4226$, corresponding to exclusion of the scale-invariant spectrum with a confidence level of $81.1\%$. These figures are highly dependent on our choice of priors: if we use a six-parameter fit with a prior of $r=0$, the likelihood of the point $n=1$ is $0.07392$, corresponding to the exclusion of the scale-invariant spectrum with a confidence level of $97.8\%$.
(In all cases, the likelihood is calculated as the marginalized posterior probability, with the best-fit point normalized to unity. The confidence level is calculated from the likelihood assuming a Gaussian error distribution.) (2) There is no
evidence for running of the spectral index of scalar perturbations. Figure
\ref{fig:run_ndn} shows likelihood contours for $dn/d \ln{k}$ relative to the
parameters $n$ and $r$, calculated at a pivot scale of $k=0.017$ to minimize
parameter degeneracies, showing that the case of a pure power-law spectrum
$dn/d \ln{k} = 0$ is fully consistent with all data sets.  (3) Potentials of
the form $V \propto \phi^p$ are consistent with the data for $p = 2$, and are
ruled out by the WMAP five-year data alone for $p = 4$. This is a stronger
conclusion that was possible with the WMAP three-year data, which was
marginally consistent with $\lambda \phi^4$ at a $95\%$ C.L. We find no
evidence for a nonzero tensor/scalar ratio, with a $95\%$ C.L. upper limit of
$r < 0.4$ for the WMAP five-year data alone, and $r < 0.35$ for WMAP in
combination with ACBAR.  Our results are consistent with the results of Peiris and Easther in Ref. \cite{Peiris:2008be}.

\begin{figure} 
\includegraphics[width=4in]{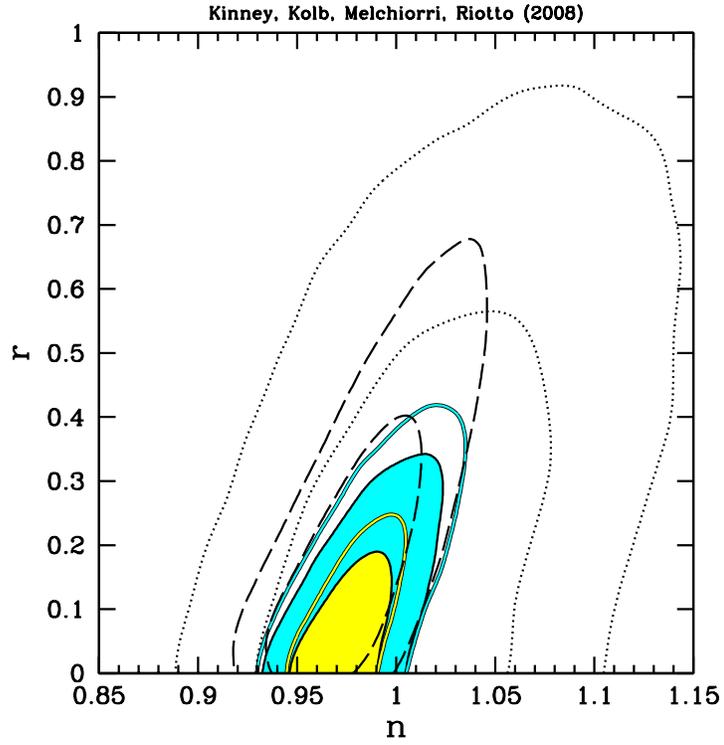}

\caption{\label{march} $68$\% and $95$\% confidence limits on  $r$ and $n$ with
a prior of $dn/d\ln{k} = 0$ for pre-WMAP datasets (see text, dotted contours), 
WMAP three-year data (dashed contours) WMAP five-year data (solid contours), and
WMAP five-year data plus ACBAR (filled contours). The pre-WMAP
limits were generated using the same parameters as the WMAP contours, 
except that the reionization optical depth was constrained to be $\tau
< 0.2$.}

\end{figure}

Finally, we illustrate the consistency of CMB determinations of $n$ and $r$
over the years in Fig.\ \ref{march}.  The pre-WMAP status is well described by
the combination of the  COBE \cite{COBE}, BOOMERANG \cite{Netterfield01},
MAXIMA \cite{Hanany00},  DASI \cite{Halverson01}, and CBI \cite{Pearson02} data
sets. This pre-WMAP analysis includes also a prior on the optical depth  $\tau
\le 0.2$.  As illustrated by the figure, higher precision data prefers $n<1$,
but $n=1$ is not strongly disfavored.

%%%%%%%%%%%%%%%%%%%%%%%%%%%%%%%%%%%%%%%%%%%%%%%%%%%%%%%%%%%%%%%%%%%%%%%%%
%%%%%%%%%%%%%%%%%%%%%%%%%%%%%%%%%%%%%%%%%%%%%%%%%%%%%%%%%%%%%%%%%%%%%%%%%
\acknowledgments
%%%%%%%%%%%%%%%%%%%%%%%%%%%%%%%%%%%%%%%%%%%%%%%%%%%%%%%%%%%%%%%%%%%%%%%%%
%%%%%%%%%%%%%%%%%%%%%%%%%%%%%%%%%%%%%%%%%%%%%%%%%%%%%%%%%%%%%%%%%%%%%%%%%

We acknowledge support provided by the Center for Computational Research at the
University at Buffalo. WHK is supported in part by the National Science
Foundation under grant NSF-PHY-0456777.  This research was supported in part by
the Department of Energy and the European Community's Research Training
Networks under contracts MRTN-CT-2004-503369, MRTN-CT-2006-035505. 
This research has been supported by ASI contract I/016/07/0 ``COFIS.''
We thank Joanna Dunkley and Eichiro Komatsu for helpful conversations.

%%%%%%%%%%%%%%%%%%%%%%%%%%%%%%%%%%%%%%%%%%%%%%%%%%%%%%%%%%%%%%%%%%%%%%%%%
%%%%%%%%%%%%%%%%%%%%%%%%%%%%%%%%%%%%%%%%%%%%%%%%%%%%%%%%%%%%%%%%%%%%%%%%%

%%%%%%%%%%%%%%%%%%%%%%%%%%%%%%%%%%%%%%%%%%%%%%%%%%%%%%%%%%%%%%%%%%%%%%%%%
%%%%%%%%%%%%%%%%%%%%%%%%%%%%%%%%%%%%%%%%%%%%%%%%%%%%%%%%%%%%%%%%%%%%%%%%%

\newpage

%%%%%%%%%%%%%%%%%%%%%%%%%%%%%%%%%%%%%%%%%%%%%%%%%%%%%%%%%%%%%%%%%%%%%%%%%
%%%%%%%%%%%%%%%%%%%%%%%%%%%%%%%%%%%%%%%%%%%%%%%%%%%%%%%%%%%%%%%%%%%%%%%%%

%%%%%%%%%%%%%%%%%%%%%%%%%%%%%%%%%%%%%%%%%%%%%%%%%%%%%%%%%%%%%%%%%%%%%%%%%
%%%%%%%%%%%%%%%%%%%%%%%%%%%%%%%%%%%%%%%%%%%%%%%%%%%%%%%%%%%%%%%%%%%%%%%%%

%%%%%%%%%%%%%%%%%%%%%%%%%%%%%%%%%%%%%%%%%%%%%%%%%%%%%%%%%%%%%%%%%%%%%%%%%
%%%%%%%%%%%%%%%%%%%%%%%%%%%%%%%%%%%%%%%%%%%%%%%%%%%%%%%%%%%%%%%%%%%%%%%%%

%%%%%%%%%%%%%%%%%%%%%%%%%%%%%%%%%%%%%%%%%%%%%%%%%%%%%%%%%%%%%%%%%%%%%%%%%
%%%%%%%%%%%%%%%%%%%%%%%%%%%%%%%%%%%%%%%%%%%%%%%%%%%%%%%%%%%%%%%%%%%%%%%%%

\end{document}